\documentclass[a4paper,12pt]{article}

\usepackage{graphicx}

\renewcommand{\baselinestretch}{1.4}

\newcommand{\pbraket}[2]%
{\ensuremath{\bigl\langle\mathsf{#1}\bigr|\mathsf{#2}\bigr\rangle}}

\newcommand{\quantph}[1]{eprint quant-ph/\linebreak[0]#1}
\makeatletter
\newcommand{\ps@first}{%
\renewcommand{\@evenhead}{}
\renewcommand{\@oddhead}{%
\hfill\raisebox{2ex}{\small To appear in Laser Physics}}
\renewcommand{\@evenfoot}{}
\renewcommand{\@oddfoot}{}
}
\makeatother

\begin{document}
\thispagestyle{first}

\begin{center}\bfseries\large
Single-loop interferometer for minimal ellipsometry%
\end{center}

\begin{center}\normalfont\normalsize
Berthold-Georg \textsc{Englert},$^1$
\textsc{Tin} Kah Ming,$^1$
\textsc{Goh} Choon Guan,$^1$\\
and \textsc{Ng} Hui Khoon$^{1,2}$
\end{center}

\begin{center}\normalfont\small
$^1$Department of Physics, National University of Singapore, Singapore 117542\\
$^2$Applied Physics Lab, DSO National Laboratories, Singapore 118230
\end{center}

\begin{center}
  (21 August 2004)
\end{center}

\begin{center}\textbf{Abstract}\\
\begin{minipage}[t]{0.9\textwidth}
We present a simple polarizing Mach-Zehnder interferometer that can be used
for optimal minimal ellipsometry: Only four intensities are measured to
determine the three Stokes parameters, and an optimal choice for
the four polarization projections can be achieved for any sufficiently small
wavelength range of interest.
\end{minipage}
\end{center}

\vfill

\begin{center}
  Dedicated to Professor Herbert Walther\\ --- grandmaster of
  optics, classical and quantum ---\\  on the occasion of his 70th birthday.
\end{center}

\newpage

The polarization properties of light --- be it emitted by a laser source,
for instance, reflected from a surface under study, or emanating from some 
sample tissue of interest --- need to be determined in many applications.  
It is, therefore, a common and frequent task in an optics laboratory to
establish the values of the three Stokes parameters that quantify the
polarization in a standard and convenient way.
The usual procedure is to measure them one by one, which is straightforward
but not very efficient.
We present here a simple interferometric setup by which one can get all three
Stokes parameters simultaneously and efficiently.

\renewcommand{\baselinestretch}{1.0}
\begin{figure}[!b]
\centerline{\includegraphics{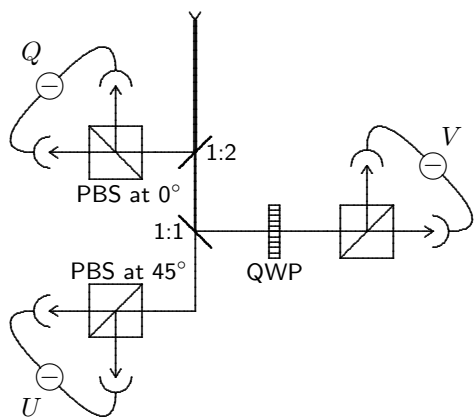}}
\caption{\label{fig:standard}\small%
Six-output setup for standard ellipsometry.
One third of the incoming light intensity is analyzed by a polarizing beam
splitter (PBS) at $0^\circ$ to establish the value of Stokes parameter $Q$
\cite{0deg}.
The remaining two thirds are distributed evenly to two more PBSs, one set at
$45^\circ$ for determining Stokes parameter $U$, the other behind a
quarter-wave plate (QWP) at $45^\circ$ for Stokes parameter $V$.~\hrulefill}
\end{figure}
\renewcommand{\baselinestretch}{1.4}

All standard \emph{ellipsometers} (or \emph{polarimeters}) are essentially
employing a setup of the kind depicted in Fig.~\ref{fig:standard}. 
In this compact design, all six intensities are measured simultaneously, but
it is, of course, also possible to carry out three consecutive
measurements of two intensities each, for which Figs.~5, 7, and 8 in
Ref.~\cite{standard} give a recent example.
One pair of detectors measures the intensities for vertical and horizontal
linear polarization, $I_\mathrm{V}$ and $I_\mathrm{H}$, and so determines
the first Stokes parameter in accordance with
\begin{equation}
  \label{eq:Q}
  Q=\frac{I_\mathrm{V}-I_\mathrm{H}}{I_\mathrm{V}+I_\mathrm{H}}\,.
\end{equation}
Another pair measures the intensities for linear polarization half-way between
horizontal and vertical, denoted by $\pm45^\circ$, yielding the second Stokes
parameter
\begin{equation}
  \label{eq:U}
  U=\frac{I_{+45}-I_{-45}}{I_{+45}+I_{-45}}\,.
\end{equation}
And the third pair measures the intensities for right-circular and
left-circular light to establish the third Stokes parameter,
\begin{equation}
  \label{eq:V}
  V=\frac{I_\mathrm{R}-I_\mathrm{L}}{I_\mathrm{R}+I_\mathrm{L}}\,.
\end{equation}
Since the inequality
\begin{equation}
  \label{eq:length}
  Q^2+U^2+V^2\leq1
\end{equation}
is necessarily obeyed, the \emph{Stokes vector}
\begin{equation}
  \label{eq:Svec}
  \vec{S}=\left(\begin{array}{c}Q\\U\\V\end{array}\right)
\end{equation}
identifies a point inside the so-called Poincar\'e sphere, 
$\bigl|\vec{S}\bigr|\leq1$.
On the surface of the sphere, we have pure polarization states, linear
polarization on the equator and circular polarization at the poles, and points
inside the sphere mark states of mixed polarization, with ``completely mixed'' 
(that is: $Q=U=V=0$) at the center of the sphere.
All of this is standard textbook wisdom.

\begin{figure}[!t]
\centerline{\includegraphics{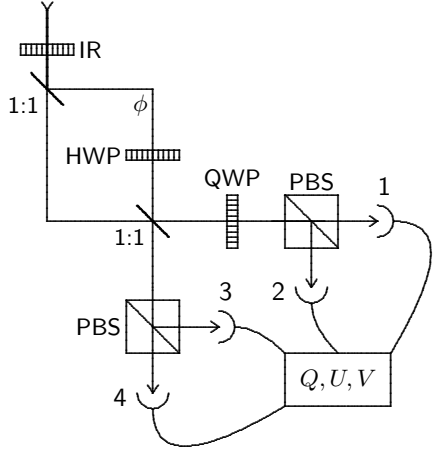}}
\caption{\label{fig:minimal}\small%
Four-output single-loop interferometer for minimal ellipsometry.
The light passes through a Mach-Zehnder interferometer that has a half-wave 
plate (HWP) at $45^\circ$ in one arm and a path-length difference that 
corresponds to a relative phase $\phi$ of 
$\mathrm{e}^{\mathrm{i}\phi}=(\sqrt{2}+\mathrm{i})/\sqrt{3}$.
The light of one output port is analyzed directly by a polarizing beam
splitter (PBS), while that emerging from the other port is first sent through
a quarter-wave plate (QWP) at $45^\circ$.
The values of the three Stokes parameters are then obtained as linear
combinations of the four relative intensities measured by the photodiodes
\textsf{1, 2, 3}, and \textsf{4}. 
The input rotator (IR) is a set of wave plates for a global unitary
polarization transformation.~\hrulefill}
\end{figure}

There are just three Stokes parameters, so that one should be able to
establish their values by measuring four intensities only, rather than six.
The interferometric setup of Fig.~\ref{fig:minimal} achieves this indeed.
The intensities $I_1$, \dots, $I_4$ measured by the four photodiodes are
related to the Stokes parameters by
\begin{eqnarray}
  \label{eq:QUV2I}
  \left.\begin{array}{c}I_1\\I_2\end{array}\right\}
&=&\frac{I}{4}\left(1-\frac{U\pm\sqrt{2}\,Q}{\sqrt{3}}\right)\,,
\nonumber\\
  \left.\begin{array}{c}I_3\\I_4\end{array}\right\}
&=&\frac{I}{4}\left(1+\frac{U\pm\sqrt{2}\,V}{\sqrt{3}}\right)\,,
\end{eqnarray}
where $I=I_1+I_2+I_3+I_4$ is the total intensity \cite{ToCome}.
Accordingly, the Stokes parameters are readily available,
\begin{eqnarray}
  \label{eq:I2QUV}
  Q&=&\sqrt{6}\,(I_2-I_1)/I\,,\nonumber\\
  U&=&\sqrt{3}\,(I_3+I_4-I_1-I_2)/I\,,\nonumber\\
  V&=&\sqrt{6}\,(I_3-I_4)/I\,.
\end{eqnarray}

\renewcommand{\baselinestretch}{1.0}
\renewcommand{\baselinestretch}{1.4}

\begin{figure}[!b]
\centerline{\includegraphics{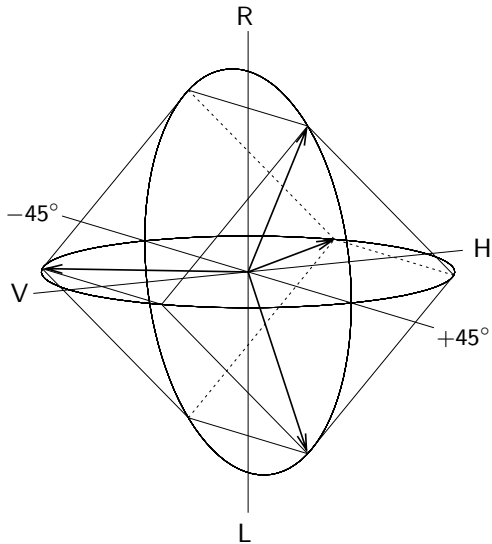}}
\caption{\label{fig:Poincare}\small%
The tetrahedron vectors of Eqs.~(\ref{eq:tetra1}) point to nonadjacent
corners of a cube that is inscribed to the Poincar\'e sphere.
Four corners of the cube, those for vectors $\vec{a}_1$ and $\vec{a}_2$ and
opposite to them, are on the equator where Stokes parameter $V$
vanishes. 
The other four corners, those for vectors  $\vec{a}_3$ and $\vec{a}_4$ and
opposite to them, are on the vertical great circle where Stokes parameter $Q$
vanishes.
On the axis \textsf{H}$\to$\textsf{V} we have the polarization states with
$U=V=0$ that can be mixed by blending horizontal and vertical polarization
only.
The other equatorial axis $\mathsf{-45^\circ\to+45^\circ}$ marks the $Q=V=0$
states that result from mixing the linear polarizations that are half-way
between horizontal and vertical.
On the vertical axis \textsf{L}$\to$\textsf{R} we have $Q=U=0$, corresponding
to polarization states that one gets when mixing left-circular with
right-circular polarization.~\hrulefill} 
\end{figure}

The relative intensities $I_j/I$ are essentially projections of the Stokes
vector onto four particular directions, 
$4I_j/I=1+\vec{a}_j\cdot\vec{S}$, that are given by
\begin{equation}
  \label{eq:tetra1}
\left.\begin{array}{c}\vec{a}_1 \\ \vec{a}_2\end{array}\right\}
=\left(\begin{array}{c}\mp\sqrt{2/3} \\ -\sqrt{1/3} \\ 0\end{array}\right)\,,
\qquad
\left.\begin{array}{c}\vec{a}_3 \\ \vec{a}_4\end{array}\right\}
=\left(\begin{array}{c}0\\ \sqrt{1/3} \\ \pm\sqrt{2/3}\end{array}\right)\,.
\end{equation}
The angle between any two of them is the same,
\begin{equation}
  \label{eq:tetra2}
  \vec{a}_j\cdot\vec{a}_k=\frac{4}{3}\delta_{jk}-\frac{1}{3}=\left\{
    \begin{array}{r@{\ \textrm{for}\ }l}
    1 & j=k\,,\\   -1/3 & j\neq k\,.
    \end{array}\right.
\end{equation}
This is to say that they realize the perfect tetrahedron geometry, which is
known to be optimal for minimal ellipsometry \cite{MQT}.
An easy way to think of these vectors is that they point from the center of a
cube to nonadjacent corners, with the cube inscribed into the Poincar\'e
sphere.
These matters are illustrated in Fig.~\ref{fig:Poincare}.

\renewcommand{\baselinestretch}{1.0}
\renewcommand{\baselinestretch}{1.4}

By a suitably chosen combination of wave plates for the unitary polarization
transformation labeled by IR in Fig.~\ref{fig:minimal}, an overall
rotation of the vector quartet (\ref{eq:tetra1}) can be performed.
This enables the experimenter to work with the tetrahedron of her choosing.

It should be clear that the setup of Fig.~\ref{fig:minimal} is not unique 
for the purpose of implementing minimal ellipsometry of this optimal kind.
For example, there is also a setup that uses polarizing beam splitters at the
entry and exit ports of the Mach-Zehnder interferometer instead of the
polarization-insensitive elements in Fig.~\ref{fig:minimal}.

Further we note that the interferometer of Fig.~\ref{fig:minimal} has a single 
loop and two output ports, whereas 
some alternative setups have two loops \cite{RenesDiss}, 
or a single loop with more output ports, among them the interferometer
of the experiment by Clarke \textit{et al.} \cite{Clarke}. 
Yet another setup has no loop at all \cite{newSetup}.

The perfect tetrahedron quartet of Eqs.~(\ref{eq:tetra1}) and
(\ref{eq:tetra2}) is realized by the setup of Fig.~\ref{fig:minimal} only if
all optical elements are just right, that is: 
the beam splitters split 1:1 for all polarizations, 
the wave plates introduce phase differences of exactly $\pi$ and $\pi/2$ 
and are precisely set at $45^\circ$, 
the path difference corresponds truly to the desired interferometer phase,
the polarizing beam splitters have ideal properties as well,
and the four photodiodes have identical efficiencies.
In practice, all these conditions can be met for a small wavelength range
only, if at all, so that distorted tetrahedrons, one for each wavelength
range, will typically be obtained in a real experiment.
Rather than Eqs.~(\ref{eq:QUV2I})--(\ref{eq:tetra2}), we then have
\begin{equation}
  \label{eq:distort1}
  \begin{array}{c}\displaystyle
  I_j=\frac{I}{4}\bigl(w_j+\vec{b}_j\cdot\vec{S}\bigr)
\quad\mbox{for\ }j=1,\dots,4 \\[2ex]\displaystyle
\mbox{with\ }\sum_{j=1}^4w_j=4
\quad\mbox{and\ }\sum_{j=1}^4\vec{b}_j=0
  \end{array}
\end{equation}
for the wavelength range in question,
where the $w_j$s determine the output intensities for
unpolarized input, and the vector quartet of the
$\vec{b}_j$s form a distorted tetrahedron \cite{calibrate}.

Even when the $w_j$s deviate much from their ideal unit value and the
distortion of the tetrahedron borders on disfigurement, the proper 
functioning as an ellipsometer is assured as long as one can solve the four
equations of (\ref{eq:distort1}) for the Stokes vector $\vec{S}$.
This is achieved by \cite{losses}
\begin{equation}
  \label{eq:distort2}
  \vec{S}=\frac{1}{4}\sum_{j=1}^4w_j\vec{c}_j
          -\frac{1}{I}\sum_{j=1}^4I_j\vec{c}_j\,,
\end{equation}
where
\begin{equation}
  \label{eq:distort3}
  \vec{c}_1=\frac{\vec{b}_2\times\vec{b}_3+\vec{b}_3\times\vec{b}_4
                  +\vec{b}_4\times\vec{b}_2}
            {\vec{b}_2\cdot(\vec{b}_3\times\vec{b}_4)}
\end{equation}
and cyclic permutations $1\to2\to3\to4\to1$ give $\vec{c}_2$, $\vec{c}_3$, 
and $\vec{c}_4$.
As a consequence, we just need that the denominator in (\ref{eq:distort3})
does not vanish, which is the basic geometrical requirement that the distorted 
tetrahedron has a nonzero volume.
But one should try to stay close to the ideal tetrahedron geometry because it
minimizes statistical errors \cite{MQT}.

In summary, we have presented a simple interferometric setup for minimal
ellipsometry. 
It consists of a Mach-Zehnder interferometer with polarization-changing
optical elements and polarization-sensitive intensity measurements at the
output ports.
The distribution of the incoming intensity to the four partial intensities
at the output is uniquely related to the polarization properties of the
incident light, and the three Stokes parameters can be inferred in a very
simple manner from the measured output intensities.
There is an ideal tetrahedron geometry, for the corresponding vectors in the
Poincar\'e sphere, but the setup is fully functional even when the actual
geometry deviates much from the ideal one. 

\begin{center}
  \textbf{Acknowledgments}
\end{center}
We wish to thank Janet Anders, Dagomir Kaszlikowski, Christian Kurtsiefer, 
Antia Lamas Linares, Jaroslav \v{R}eh\'{a}\v{c}ek, and Gregor Weihs 
for valuable discussions.
We gratefully acknowledge the financial support from 
Temasek Grant WBS: R-144-000-071-305.
Ng H.K. would also like to thank the Defence Science \&
Technology Agency (DSTA) of Singapore for their financial support.

\renewcommand{\refname}{\centering\normalsize\textbf{Notes and references}}
\renewcommand{\baselinestretch}{1.0}\normalfont


\begin{thebibliography}{9}
\bibitem{ToCome}
A detailed description of setups like the one in Fig.~\ref{fig:minimal} will
be given elsewhere \cite{newSetup}.
We are content here with a brief account and a statement of the most important 
facts.
\bibitem{calibrate}
There are altogether $12$ real parameters that specify the four $w_j$s
and the four $\vec{b}_j$s.
They can be determined experimentally by measuring $I_1$, \dots, $I_4$ for
four suitably chosen, known polarizations of the input light.
In this sense, then, the setup is ``self-calibrating.''
\bibitem{losses}
Equations (\ref{eq:distort2}) and (\ref{eq:distort3}) apply if there are no
polarization-dependent losses.
\end{thebibliography}
\end{document}